# A novel LMI-based Method for Robust Stabilization of Fractional-order Interval Systems with $1 \leq \alpha < 2$

Pouya Badri[1], Mahdi Sojoodi[1]

[1]Advanced Control Systems Laboratory, School of Electrical and Computer Engineering, Tarbiat Modares University, Tehran, Iran.

**Abstract:** This paper deals with the problem of robust dynamic output feedback stabilization of interval fractional-order linear time invariant (FO-LTI) systems with the fractional order $1 \leq \alpha < 2$. In this study, a new formulation based on the null-space analysis of the system matrices is proposed using linear matrix inequalities (LMIs). The applied uncertain model is the most complete model of linear interval systems, in which all of the systems matrices are interval matrices. A robust dynamic output feedback controller is designed that asymptotically stabilizes the interval FO-LTI system, where no limiting constraint is assumed on the state space matrices of the uncertain system. Eventually, a numerical example with simulations is presented to demonstrate the effectiveness and correctness of the theoretical results.

## 1. Introduction

In recent years the study of fractional-order control systems has drawn increasing interest and attention of physicists and engineers in view of applications [1–4]. It has been proved that, systems having responses with long memory transients can more concisely and actually be described by fractional-order models. Moreover, a large number of natural and biological systems have inherent properties that can be better described by fractional-order models [6–9]. Therefore, designing controllers for fractional-order systems are among the most interesting problems in the literature.

Stability is the primary objective to be accomplished in most of control systems, and because of uncertain models caused by neglected dynamics, uncertain physical parameters, parametric variations in time, and so on, robust stability and stabilization problems became a basic issue for all control systems as well as fractional-order systems [10]–[13]. It should be noted that fractional-order derivatives are nonlocal and have weakly singular kernels and the stability analysis of FO systems is more complicated than that of integer-order systems [9].

Robust stability and stabilization analysis on fractional-order systems were investigated in [10]–[16]. Stability and stabilization of fractional-order interval systems are investigated in [10] where necessary and sufficient conditions for robust stability and stabilization of such systems are presented in the form of LMI. Furthermore, In [11] the robust stability and stabilization of fractional-order linear systems with positive real uncertainty are investigated, in which the existence conditions and design procedures of the static state feedback controller, static output feedback controller and observer-based controller for asymptotically stabilizing of such systems are presented with the constraint on the output matrix to be of full-row rank. In [16] the problem of robust stability for fractional-order linear systems with polytopic uncertainties is investigated where based on the Kronecker product and µ-analysis, stable condition is established. Numerical methods for robust stability analysis of closed-loop control systems with parameter uncertainty was presented in [17] based on scan sampling of interval characteristic polynomials from the hypercube of parameter space.

In the most of listed studies, state feedback controller is used, where all individual states are needed. but, in most practical applications it is not possible to measure all states due to economic issues or physical constraints, where employing output feedback control is useful [19,20]. It is noteworthy that dynamic feedback controller has more effective control performances, more flexibility, and more degrees of freedom in achieving control objectives, in comparison with the static controller [21].

Motivated by this observation, this study aims at solving the problem of robust stabilization of fractional-order linear interval systems with the fractional order $1 \leq \alpha < 2$, using a dynamic output feedback controller with a predetermined order based on the null-space analysis of the system matrices in terms of linear matrix inequalities (LMIs). The contributions of this article can be summarized as follows:

- The most complete uncertain interval model including interval output matrix $C$ is considered unlike previous works.
- A new formulation based on the null-space analysis of the system matrices is presented for robust stabilization of interval FO-LTI systems for the first time.
- Utilizing dynamic controller more performance efficiency and degrees of freedom are achieved and not all of the pseudo-states of the uncertain FO-LTI system is needed, thanks to the output feedback scheme.
- The LMI-based approach of developing robust stabilizing control, which can be numerically solved by convex programming problem, is preserved in spite of the complexity of assuming the most complete model of uncertain system, including interval output matrix $C$ and the most complete model of linear controller, including direct feedthrough parameter.

This article is structured as follows. In section 2, some preliminaries about interval uncertainty, null-space of a matrix, and fractional-order calculus together with the problem formulation are presented.

Robust stabilizing conditions of uncertain fractional-order systems with interval uncertainty via fixed-order dynamic output feedback controller alongside with the design procedures of the corresponding controllers are derived in Section 3. Simulation results are presented in Section 4, while conclusions are drawn in Section 5.

**Notations**: In this paper $A \otimes B$ denotes the kronecker product of matrices $A$ and $B$ and by $M^T$, $\bar{M}$ and $M^*$, we denote the transpose, the conjugate, and the transpose conjugate of $M$, resepectively. The conjugate of the scalar number $z$ is represented by $\bar{z}$ and $Sym(M)$ denotes $M + M^*$. The notation ● is the symmetric component symbol in matrix and ↑ is the symbol of pseudo inverse. The notations **0** and $I$ denote the zero and identity matrices with appropriate dimensions and $i$ stands for the imaginary unit.

## 2. Preliminaries and problem formulation

In this section, some basic concepts and lemmas of fractional-order calculus and interval uncertainty are presented.

Consider the following uncertain FO-LTI system:

$$D^\alpha x(t) = Ax(t) + Bu(t), 1 \leq \alpha < 2 \tag{1}$$
$$y(t) = Cx(t)$$

in which $x \in R^n$ denotes the pseudo-state vector, $u \in R^l$ is the control input, and $y \in R^m$ is the output vector. Furthermore, $A \in R^{n \times n}$, $B \in R^{n \times l}$, and $C \in R^{m \times n}$ are interval uncertain matrices as follows

$$A \in A_I = [\underline{A}, \overline{A}] = \{[a_{ij}]: \underline{a}_{ij} \leq a_{ij} \leq \overline{a}_{ij}, 1 \leq i,j \leq n\}, \tag{2}$$

$$B \in B_I = [\underline{B}, \overline{B}] = \{[b_{ij}]: \underline{b}_{ij} \leq b_{ij} \leq \overline{b}_{ij}, 1 \leq i \leq n, 1 \leq j \leq l\}, \tag{3}$$

$$C \in C_I = [\underline{C}, \overline{C}] = \{[c_{ij}]: \underline{c}_{ij} \leq c_{ij} \leq \overline{c}_{ij}, 1 \leq i \leq m, 1 \leq j \leq n\}, \tag{4}$$

where $\underline{A} = [\underline{a}_{ij}]_{n \times n}$ and $\overline{A} = [\overline{a}_{ij}]_{n \times n}$ satisfy $\underline{a}_{ij} \leq \overline{a}_{ij}$ for all $1 \leq i,j \leq n$, $\underline{B} = [\underline{b}_{ij}]_{n \times l}$ and $\overline{B} = [\overline{b}_{ij}]_{n \times l}$ satisfy $\underline{b}_{ij} \leq \overline{b}_{ij}$ for all $1 \leq i \leq n, 1 \leq j \leq l$, and $\underline{C} = [\underline{c}_{ij}]_{m \times n}$ and $\overline{C} = [\overline{c}_{ij}]_{m \times n}$ satisfy $\underline{c}_{ij} \leq \overline{c}_{ij}$ for all $1 \leq i \leq m, 1 \leq j \leq n$.

In this paper, the following Caputo definition for fractional derivatives of order α of function $f(t)$ is adopted since initial values of classical integer-order derivatives with clear physical interpretations are utilizable using the Laplace transform of the Caputo derivative [22]:

$$^C_a D^\alpha_t f(t) = \frac{1}{\Gamma(m-\alpha)} \int_a^t (t-\tau)^{m-a-1} \left(\frac{d}{d\tau}\right)^m f(\tau)d\tau$$

where $\Gamma(\cdot)$ is Gamma function defined by $\Gamma(\epsilon) = \int_0^\infty e^{-t} t^{\epsilon-1} dt$ and $m$ is the smallest integer that is equal to or greater than $\alpha$.

The following notations are introduced in order to deal with the interval uncertainties

$$A_0 = 1/2(\underline{A} + \overline{A}), \Delta A = 1/2(\overline{A} - \underline{A}) = \{\gamma_{ij}\}_{n \times n} \tag{5}$$

$$B_0 = 1/2(\underline{B} + \overline{B}), \Delta B = 1/2(\overline{B} - \underline{B}) = \{\beta_{ij}\}_{n \times l} \tag{6}$$

$$C_0 = 1/2(\underline{C} + \overline{C}), \Delta C = 1/2(\overline{C} - \underline{C}) = \{\varepsilon_{ij}\}_{m \times n} \tag{7}$$

It is obvious that all elements of $\Delta A$, $\Delta B$, and $\Delta C$ are nonnegative, so the following matrices can be defined

$$M_A = [\sqrt{\gamma_{11}} e^n_1 \quad \ldots \quad \sqrt{\gamma_{1n}} e^n_1 \quad \ldots \quad \sqrt{\gamma_{n1}} e^n_n \quad \ldots \quad \sqrt{\gamma_{nn}} e^n_n]_{n \times n^2}, \tag{8}$$

$$R_A = [\sqrt{\gamma_{11}} e^n_1 \quad \ldots \quad \sqrt{\gamma_{1n}} e^n_n \ldots \quad \sqrt{\gamma_{n1}} e^n_1 \quad \ldots \quad \sqrt{\gamma_{nn}} e^n_n]^T_{n^2 \times n}, \tag{9}$$

$$M_B = [\sqrt{\beta_{11}} e^n_1 \quad \ldots \quad \sqrt{\beta_{1l}} e^n_1 \quad \ldots \sqrt{\beta_{n1}} e^n_n \quad \ldots \quad \sqrt{\beta_{nl}} e^n_n]_{n \times nl}, \tag{10}$$

$$R_B = [\sqrt{\beta_{11}} e^l_1 \quad \ldots \quad \sqrt{\beta_{1l}} e^l_l \quad \ldots \sqrt{\beta_{n1}} e^l_1 \quad \ldots \quad \sqrt{\beta_{nl}} e^l_l]^T_{nl \times l}, \tag{11}$$

$$M_C = [\sqrt{\varepsilon_{11}} e^m_1 \quad \ldots \quad \sqrt{\varepsilon_{1n}} e^m_1 \quad \ldots \sqrt{\varepsilon_{m1}} e^m_m \quad \ldots \quad \sqrt{\varepsilon_{mn}} e^m_m]_{m \times mn}, \tag{12}$$

$$R_C = [\sqrt{\varepsilon_{11}} e^n_1 \quad \ldots \quad \sqrt{\varepsilon_{1n}} e^n_n \quad \ldots \sqrt{\varepsilon_{m1}} e^n_1 \quad \ldots \quad \sqrt{\varepsilon_{mn}} e^n_n]^T_{mn \times n}, \tag{13}$$

where $e^n_k \in R^n$, $e^l_k \in R^l$, and $e^m_k \in R^m$ are column vectors with the k-th element being 1 and all the others being 0. Also, we have

$$H_A = \{diag(\delta_{11}, \ldots, \delta_{1n}, \ldots, \delta_{n1}, \ldots, \delta_{nn}) \in R^{n^2 \times n^2}, |\delta_{ij}| \leq 1, i,j\ 1, \ldots, n\}, \tag{14}$$

$$H_B = \{diag(\eta_{11}, \ldots, \eta_{1l}, \ldots, \eta_{n1}, \ldots, \eta_{nl}) \in R^{(nl) \times (nl)}, |\eta_{ij}| \leq 1, i = 1, \ldots, n, j = 1, \ldots, l\}, \tag{15}$$

$$H_C = \{diag(\upsilon_{11}, \ldots, \upsilon_{1n}, \ldots, \upsilon_{m1}, \ldots, \upsilon_{mn}) \in R^{(mn) \times (mn)}, |\upsilon_{ij}| \leq 1, i = 1, \ldots, m, j = 1, \ldots, n\}. \tag{16}$$

In order to study the stability of fractional-order systems and obtain main results the following lemmas are required.

**Lemma 1** [10]: Let

$$A_J = \{A = A_0 + M_A F_A R_A | F_A \in H_A\}, B_J = \{B = B_0 + M_B F_B R_B | F_B \in H_B\}, \quad (17)$$
$$C_J = \{C = C_0 + M_C F_C R_C | F_C \in H_C\}.$$

Then $A_I = A_J$, $B_I = B_J$, and $C_I = C_J$.

**Lemma 2** [23]: Let $A \in \mathcal{R}^{n \times n}$, $1 \leq \alpha < 2$ and $\theta = \pi - \alpha\pi/2$. The fractional-order system $D^\alpha x(t) = Ax(t)$ is asymptotically stable if and only if there exists a positive definite matrix $X \in \mathcal{R}^{n \times n}$ such that

$$\begin{bmatrix} (A^T X + XA)\sin\theta & (XA - A^T X)\cos\theta \\ \bullet & (A^T X + XA)\sin\theta \end{bmatrix} < 0, \quad (18)$$

defining

$$\Theta = \begin{bmatrix} \sin\theta & -\cos\theta \\ \cos\theta & \sin\theta \end{bmatrix}, \quad (19)$$

and with this in mind that $A$ is similar to $A^T$, inequality (15) can be expressed as follows

$$Sym\{\Theta \otimes (AX)\} < 0. \quad (20)$$

**Lemma 3** [10]: For any matrices $X$ and $Y$ with appropriate dimensions, we have

$$X^T Y + Y^T X \leq \eta X^T X + (1/\eta) Y^T Y \text{ for any } \eta > 0. \quad (21)$$

**Definition 1.** [24] The kernel (also known as null-space) of a linear map $A : V \to W$ between two vector spaces $V$ and $W$, is as follows

$$\ker A = \{v \in V | Av = 0\}, \quad (22)$$

where 0 denotes the zero vector in $W$. And the image of $A : V \to W$ is defined as

$$\text{Im } A = \{w \in W | \exists v \in V: Av = w\}. \quad (23)$$

**Lemma 4** [25]: For any matrices $P$ and $Q$, together with symmetric matrix $H$, there exists a matrix $J$ such that

$$H + P^T J^T Q + Q^T J P < 0, \quad (24)$$

if and only if

$$N_P^T H N_P < 0, \quad N_Q^T H N_Q < 0, \quad (25)$$

in which $N_P$ and $N_Q$ are full-rank matrices in a way that

$$\text{Im } N_P = \ker P, \quad \text{Im } N_Q = \ker Q. \quad (26)$$

**Lemma 5** [25]: For any symmetric positive definite matrices $X, Y \in R^{n \times n}$ there exist $X_2, Y_2 \in R^{n \times n_c}$, and $X_3, Y_3 \in R^{n_c \times n_c}$ matrices that satisfy the conditions

$$\begin{bmatrix} X & X_2 \\ X_2^T & X_3 \end{bmatrix} \geq 0, \begin{bmatrix} X & X_2 \\ X_2^T & X_3 \end{bmatrix}^{-1} = \begin{bmatrix} Y & Y_2 \\ Y_2^T & Y_3 \end{bmatrix}, \quad (27)$$

if and only if

$$\begin{bmatrix} X & I \\ I & Y \end{bmatrix} \geq 0, Rank\left(\begin{bmatrix} X & I \\ I & Y \end{bmatrix}\right) \leq n + n_c. \quad (28)$$

## 3. Main results

The main objective of this paper is to design a robust dynamic output feedback controller that asymptotically stabilizes the interval FO-LTI system (1) in terms of linear matrix inequalities (LMIs). Therefore, the following dynamic output feedback controller is presented

$$D^\alpha x_C(t) = A_C x_C(t) + B_C y(t), \quad 1 \leq \alpha < 2 \quad (29)$$
$$u(t) = C_C x_C(t) + D_C y(t),$$

with $x_c \in \mathcal{R}^{n_c}$, in which $n_c$ is the arbitrary order of the controller and $A_C, B_C, C_C,$ and $D_C$ are corresponding matrices to be designed.

The resulted closed-loop augmented FO-LTI system using (1) and (29) is as follows

$$D^\alpha x_{Cl}(t) = A_{Cl} x_{Cl}(t), \quad 1 \leq \alpha < 2 \tag{30}$$

where

$$x_{Cl}(t) = \begin{bmatrix} x(t) \\ x_C(t) \end{bmatrix}, \quad A_{Cl} = \begin{bmatrix} A + BD_C C & BC_C \\ B_C C & A_C \end{bmatrix}. \tag{31}$$

**Theorem 1:** Considering closed-loop system in (30) with $1 \leq \alpha < 2$, a positive definite symmetric matrix $X_{Cl} = X_{Cl}^T \in \mathcal{R}^{(n+n_c) \times (n+n_c)}$ with $n_c \geq n$, and real scalar constants $\eta_i > 0, i = 1, \ldots, 7$ exist in a way that following LMI constrains become feasible

$$\begin{bmatrix} \sigma_i + \eta_i M_i M_i^T & R_i^T \\ \bullet & -\eta_i I \end{bmatrix} < 0, i = 1, 2, \begin{bmatrix} X & I \\ I & Y \end{bmatrix} \geq 0, \begin{bmatrix} \sigma_3 + \sum_{i=3}^{7} \eta_i M_i M_i^T & R_3^T & R_4^T & R_5^T & R_6^T & R_7^T \\ \bullet & -\eta_3 I & 0 & 0 & 0 & 0 \\ \bullet & \bullet & -\eta_4 I & 0 & 0 & 0 \\ \bullet & \bullet & \bullet & -\eta_5 I & 0 & 0 \\ \bullet & \bullet & \bullet & \bullet & -\eta_6 I & 0 \\ \bullet & \bullet & \bullet & \bullet & \bullet & -\eta_7 I \end{bmatrix} < 0, \tag{32}$$

in which

$$\sigma_1 = \begin{bmatrix} N_C^T(A_0 Y + YA_0^T)N_C(\sin\theta)^{-1} & N_C^T(YA_0^T - A_0 Y)N_C(\cos\theta)^{-1} \\ N_C^T(A_0 Y - YA_0^T)N_C(\cos\theta)^{-1} & N_C^T(A_0 Y + YA_0^T)N_C(\sin\theta)^{-1} \end{bmatrix},$$

$$\sigma_2 = \begin{bmatrix} N_O^T(A_0^T X + XA_0)N_O \sin\theta & N_O^T(XA_0 - A_0^T X)N_O \cos\theta \\ N_O^T(A_0^T X - XA_0)N_O \cos\theta & N_O^T(A_0^T X + XA_0)N_O \sin\theta \end{bmatrix},$$

$$\sigma_3 = \begin{bmatrix} (\mathbb{A}_0^T X_{Cl} + X_{Cl} \mathbb{A}_0) \sin\theta & (X_{Cl} \mathbb{A}_0 - \mathbb{A}_0^T X_{Cl}) \cos\theta \\ (\mathbb{A}_0^T X_{Cl} - X_{Cl} \mathbb{A}_0) \cos\theta & (\mathbb{A}_0^T X_{Cl} + X_{Cl} \mathbb{A}_0) \sin\theta \end{bmatrix},$$

$$R_1 = \begin{bmatrix} R_A Y N_C (\sin\theta)^{-1} & -R_A Y N_C (\cos\theta)^{-1} \\ R_A Y N_C (\cos\theta)^{-1} & R_A Y N_C (\sin\theta)^{-1} \end{bmatrix}, R_2 = I_2 \otimes M_A X N_O, R_3 = \begin{bmatrix} M_A R_A \sin\theta & 0 & M_A R_A \cos\theta & 0 \\ 0 & 0 & 0 & 0 \\ -M_A R_A \cos\theta & 0 & M_A R_A \sin\theta & 0 \\ 0 & 0 & 0 & 0 \end{bmatrix},$$

$$R_4 = R_5 = \begin{bmatrix} K\mathbb{C}_0 & 0 \\ 0 & K\mathbb{C}_0 \end{bmatrix}, R_6 = R_7 = \begin{bmatrix} K & 0 \\ 0 & K \end{bmatrix} \begin{bmatrix} 0 & 0 & 0 & 0 \\ M_C R_C & 0 & 0 & 0 \\ 0 & 0 & 0 & 0 \\ 0 & 0 & M_C R_C & 0 \end{bmatrix}, \tag{33}$$

$$M_1 = I_2 \otimes N_C^T M_A, M_2 = \begin{bmatrix} R_A N_O \sin\theta & R_A N_O \cos\theta \\ -R_A N_O \cos\theta & R_A N_O \sin\theta \end{bmatrix}^T, M_3 = \begin{bmatrix} X_{Cl} & 0 \\ 0 & X_{Cl} \end{bmatrix},$$

$$M_4 = M_6 = \begin{bmatrix} X_{Cl} \mathbb{B}_0 \sin\theta & X_{Cl} \mathbb{B}_0 \cos\theta \\ -X_{Cl} \mathbb{B}_0 \cos\theta & X_{Cl} \mathbb{B}_0 \sin\theta \end{bmatrix}, M_5 = M_7 = \begin{bmatrix} X_{Cl} & 0 \\ 0 & X_{Cl} \end{bmatrix} \begin{bmatrix} 0 & M_B R_B \sin\theta & 0 & M_B R_B \cos\theta \\ 0 & M_B R_B \sin\theta & 0 & M_B R_B \cos\theta \\ 0 & -M_B R_B \cos\theta & 0 & M_B R_B \sin\theta \\ 0 & -M_B R_B \cos\theta & 0 & M_B R_B \sin\theta \end{bmatrix},$$

where $\theta = \pi - \alpha \pi / 2$ then, the obtained dynamic output feedback controller parameters of (29), make the closed-loop system in (30) asymptotically stable.

**Proof.** It follows from Lemma 3 that the uncertain fractional-order closed-loop system (30) with $1 < \alpha \leq 2$ is asymptotically stable if there exists a positive definite matrix $X_{Cl} = X_{Cl}^T$, $X_{Cl} \in \mathcal{R}^{(n+n_c) \times (n+n_c)}$ such that

$$\begin{bmatrix} (A_{Cl}^T X_{Cl} + X_{Cl} A_{Cl}) \sin\theta & (X_{Cl} A_{Cl} - A_{Cl}^T X_{Cl}) \cos\theta \\ (A_{Cl}^T X_{Cl} - X_{Cl} A_{Cl}) \cos\theta & (A_{Cl}^T X_{Cl} + X_{Cl} A_{Cl}) \sin\theta \end{bmatrix} = \begin{bmatrix} (\mathbb{A}^T X_{Cl} + X_{Cl} \mathbb{A}) \sin\theta & (X_{Cl} \mathbb{A} - \mathbb{A}^T X_{Cl}) \cos\theta \\ (\mathbb{A}^T X_{Cl} - X_{Cl} \mathbb{A}) \cos\theta & (\mathbb{A}^T X_{Cl} + X_{Cl} \mathbb{A}) \sin\theta \end{bmatrix} \tag{34}$$

$$+ \begin{bmatrix} X_{Cl} \mathbb{B} \sin\theta & X_{Cl} \mathbb{B} \cos\theta \\ X_{Cl} \mathbb{B} \cos\theta & X_{Cl} \mathbb{B} \sin\theta \end{bmatrix} \begin{bmatrix} K & 0 \\ 0 & K \end{bmatrix} \begin{bmatrix} \mathbb{C} & 0 \\ 0 & \mathbb{C} \end{bmatrix} + \left( \begin{bmatrix} X_{Cl} \mathbb{B} \sin\theta & X_{Cl} \mathbb{B} \cos\theta \\ X_{Cl} \mathbb{B} \cos\theta & X_{Cl} \mathbb{B} \sin\theta \end{bmatrix} \begin{bmatrix} K & 0 \\ 0 & K \end{bmatrix} \begin{bmatrix} \mathbb{C} & 0 \\ 0 & \mathbb{C} \end{bmatrix} \right)^T < 0,$$

where

$$\mathbb{A} = \begin{bmatrix} A & 0 \\ 0 & 0 \end{bmatrix}, \mathbb{B} = \begin{bmatrix} 0 & B \\ I & 0 \end{bmatrix}, \mathbb{C} = \begin{bmatrix} 0 & I \\ C & 0 \end{bmatrix}, K = \begin{bmatrix} A_C & B_C \\ C_C & D_C \end{bmatrix}. \tag{35}$$

By defining $P_{X_{Cl}}, H_{X_{Cl}}, Q,$ and $\widetilde{K}$ matrices as follows

$$P_{X_{Cl}} = \begin{bmatrix} X_{Cl} \mathbb{B} \sin\theta & X_{Cl} \mathbb{B} \cos\theta \\ X_{Cl} \mathbb{B} \cos\theta & X_{Cl} \mathbb{B} \sin\theta \end{bmatrix}^T, \quad H_{X_{Cl}} = \begin{bmatrix} (\mathbb{A}^T X_{Cl} + X_{Cl} \mathbb{A}) \sin\theta & (X_{Cl} \mathbb{A} - \mathbb{A}^T X_{Cl}) \cos\theta \\ (\mathbb{A}^T X_{Cl} - X_{Cl} \mathbb{A}) \cos\theta & (\mathbb{A}^T X_{Cl} + X_{Cl} \mathbb{A}) \sin\theta \end{bmatrix}, \tag{36}$$

$$Q = \begin{bmatrix} \mathbb{C} & 0 \\ 0 & \mathbb{C} \end{bmatrix}, \widetilde{K} = \begin{bmatrix} K & 0 \\ 0 & K \end{bmatrix},$$

inequality (34) can be rewritten as

$$H_{X_{Cl}} + Q^T \widetilde{K}^T P_{X_{Cl}} + P_{X_{Cl}}^T \widetilde{K} Q < 0. \tag{37}$$

According to Lemma 5 inequality (37) is equivalent to

$$N_{P_{X_{Cl}}}^T H_{X_{Cl}} N_{P_{X_{Cl}}} < 0, \quad N_Q^T H_{X_{Cl}} N_Q < 0, \tag{38}$$

in which $\text{Im } N_{P_{X_{Cl}}} = \ker P_{X_{Cl}}$. Both inequalities in (38) are not linear due to various multiplications of variables. Thus, the following matrices are defined

$$T_{X_{Cl}} = \begin{bmatrix} (\mathbb{A}X_{Cl}^{-1} + X_{Cl}^{-1}\mathbb{A}^T)\sin\theta & (-X_{Cl}^{-1}\mathbb{A}^T + \mathbb{A}X_{Cl}^{-1})\cos\theta \\ (X_{Cl}^{-1}\mathbb{A}^T - \mathbb{A}X_{Cl}^{-1})\cos\theta & (\mathbb{A}X_{Cl}^{-1} + X_{Cl}^{-1}\mathbb{A}^T)\sin\theta \end{bmatrix}, P = \begin{bmatrix} 1 & -1 \\ 1 & 1 \end{bmatrix} \otimes \underline{B}^T, S = \begin{bmatrix} X_{Cl}\sin\theta & \mathbf{0} \\ \mathbf{0} & X_{Cl}\cos\theta \end{bmatrix}. \tag{39}$$

Since $P_{X_{Cl}} = PS$, it can be concluded that

$$N_{P_{X_{Cl}}} = S^{-1} N_P, \tag{40}$$

in which $\text{Im } N_P = \ker P$. Substituting (40) in the first inequality of (38) yields to

$$N_P^T (S^{-1})^T H_{X_{Cl}} S^{-1} N_P < 0, \tag{41}$$

which is equivalent to $N_P^T T_{X_{Cl}} N_P < 0$ according to definition of $H_{X_{Cl}}$ and $T_{X_{Cl}}$ in (36) and (39) respectively. Therefore, stability conditions in (38) can be rewritten as

$$N_P^T T_{X_{Cl}} N_P < 0, \quad N_Q^T H_{X_{Cl}} N_Q < 0. \tag{42}$$

The first and second inequalities of (42) are LMI conditions in $X_{Cl}^{-1}$ and $X_{Cl}$ respectively, which causes (42) not to be linear in $X_{Cl}$ in total. In order to solve this problem, $X_{Cl}$ and $X_{Cl}^{-1}$ matrices are supposed to have following structures

$$X_{Cl} = \begin{bmatrix} X & X_2 \\ X_2^T & X_3 \end{bmatrix}, \quad X_{Cl}^{-1} = \begin{bmatrix} Y & Y_2 \\ Y_2^T & Y_3 \end{bmatrix}, \tag{43}$$

in which $X_{Cl} \in R^{(n+n_C) \times (n+n_C)}$ is a positive definite symmetric matrix and submatrices $X$ and $Y$ are of $n \times n$ dimension.

By respectively substituting $\mathbb{A}$ and $X_{Cl}^{-1}$ from (35) and (43) in (39), matrix $T_{X_{Cl}}$ can be rewritten as follows

$$T_{X_{Cl}} = \begin{bmatrix} (AY + YA^T)\sin\theta & AY_2 \sin\theta & (AY - YA^T)\cos\theta & AY_2 \cos\theta \\ Y_2^T A^T \sin\theta & 0 & -Y_2^T A^T \cos\theta & 0 \\ (-AY + YA^T)\cos\theta & AY_2 \cos\theta & (AY + YA^T)\sin\theta & AY_2 \sin\theta \\ Y_2^T A^T \cos\theta & 0 & Y_2^T A^T \sin\theta & 0 \end{bmatrix}. \tag{44}$$

Moreover, by substituting $\underline{B}$ from (35) in (39), matrix $P$ can be rewritten as

$$P = \begin{bmatrix} 1 & -1 \\ 1 & 1 \end{bmatrix} \otimes \begin{bmatrix} \mathbf{0} & I \\ B^T & \mathbf{0} \end{bmatrix}. \tag{45}$$

Since $\text{Im } N_P = \ker P$, and according to the Definition 1 one can write the following equations

$$\text{Im } N_P = \ker P, \ker P = \{x | Px = 0\}, \text{Im } N_P = \{y | N_P z = y\} \tag{46}$$

from which it can be concluded that

$$N_P z = x \implies P N_P z = P x = 0 \implies P N_P z = 0, \exists z. \tag{47}$$

A possible solution for (46) would be $PN_P = 0$ which leads to $N_P = \ker P$. Therefore, $\ker P$ is calculated as below

$$\begin{bmatrix} 0 & I & 0 & -I \\ B^T & 0 & -B^T & 0 \\ 0 & I & 0 & I \\ B^T & 0 & B^T & 0 \end{bmatrix} \begin{bmatrix} v_{11} & v_{12} & v_{13} & v_{14} \\ v_{21} & v_{22} & v_{23} & v_{24} \\ v_{31} & v_{32} & v_{33} & v_{34} \\ v_{41} & v_{42} & v_{43} & v_{44} \end{bmatrix} = \mathbf{0} \implies B^T(v_{1i} - v_{3i}) = 0, i = 1, \dots, 4, \tag{48}$$

in which $N_P = (v_{ij}), i, j = 1, \dots, 4$. According to (48), a possible solution for $N_P$ is as follows

$$N_P = \begin{bmatrix} N_C & 0 & 0 & 0 \\ 0 & 0 & 0 & 0 \\ 0 & -N_C & 0 & 0 \\ 0 & 0 & 0 & 0 \end{bmatrix}, \quad (49)$$

with $N_C$ in the null-space of $B^T$, i.e. $B^T N_C = 0$. Since the second and forth rows of $N_P$ is zero, the second and forth rows and columns of matrix $T_{X_{Cl}}$ have no effect on the first inequality of (42), which can be rewritten as

$$\begin{bmatrix} N_C^T(AY+YA^T)N_C(\sin\theta)^{-1} & N_C^T(YA^T-AY)N_C(\cos\theta)^{-1} \\ N_C^T(AY-YA^T)N_C(\cos\theta)^{-1} & N_C^T(AY+YA^T)N_C(\sin\theta)^{-1} \end{bmatrix} < 0. \quad (50)$$

By some manipulation the same conclusion can be drawn for the second inequality of (42), where $N_Q = \ker Q$ is obtained in a similar way as follows

$$N_Q = \begin{bmatrix} N_O & 0 & 0 & 0 \\ 0 & 0 & 0 & 0 \\ 0 & -N_O & 0 & 0 \\ 0 & 0 & 0 & 0 \end{bmatrix}, \quad (51)$$

in which $N_O$ is in the null-space of $C$, i.e. $CN_O = 0$. Since the second and forth rows of $N_Q$ is zero, the second and forth rows and columns of matrix $H_{X_{Cl}}$ have no effect on the first inequality of (42), which can be rewritten

$$\begin{bmatrix} N_O^T(A^T X + XA)N_O \sin\theta & N_O^T(XA - A^T X)N_O \cos\theta \\ N_O^T(A^T X - XA)N_O \sin\theta & N_O^T(A^T X + XA)N_O \sin\theta \end{bmatrix} < 0. \quad (52)$$

Lemma 5 states the conditions under which according to (43) symmetric positive definite matrix $X_{Cl}$ and its inverse $X_{Cl}^{-1}$ can be constructed using $Y$ and $X$ matrices obtained from (50) and (52) LMIs respectively. Therefore, the first constraint of Lemma 5 in (28), which is an LMI condition, should be added into the stabilization constraints and the second rank constraint in (28) is satisfied by choosing $n_C \geq n$. Taking into account the uncertainties in (2)-(4), LMI inequalities in (50) and (52) can be rewritten as follows

$$\begin{bmatrix} N_C^T(AY+YA^T)N_C(\sin\theta)^{-1} & N_C^T(YA^T-AY)N_C(\cos\theta)^{-1} \\ N_C^T(AY-YA^T)N_C(\cos\theta)^{-1} & N_C^T(AY+YA^T)N_C(\sin\theta)^{-1} \end{bmatrix} =$$
$$\begin{bmatrix} N_C^T(A_0 Y+YA_0^T)N_C(\sin\theta)^{-1} & N_C^T(YA_0^T-A_0 Y)N_C(\cos\theta)^{-1} \\ N_C^T(A_0 Y-YA_0^T)N_C(\cos\theta)^{-1} & N_C^T(A_0 Y+YA_0^T)N_C(\sin\theta)^{-1} \end{bmatrix} +$$
$$Sym \left\{ \begin{bmatrix} N_C^T M_A \Delta_A R_A Y N_C(\sin\theta)^{-1} & -N_C^T M_A \Delta_A R_A Y N_C(\cos\theta)^{-1} \\ N_C^T M_A \Delta_A R_A Y N_C(\cos\theta)^{-1} & N_C^T M_A \Delta_A R_A Y N_C(\sin\theta)^{-1} \end{bmatrix} \right\} < 0, \quad (53)$$

$$\begin{bmatrix} N_O^T(A^T X + XA)N_O \sin\theta & N_O^T(XA - A^T X)N_O \cos\theta \\ N_O^T(A^T X - XA)N_O \sin\theta & N_O^T(A^T X + XA)N_O \sin\theta \end{bmatrix} =$$
$$\begin{bmatrix} N_O^T(A_0^T X + XA_0)N_O \sin\theta & N_O^T(XA - A_0^T X)N_O \cos\theta \\ N_O^T(A_0^T X - XA_0)N_O \sin\theta & N_O^T(A_0^T X + XA_0)N_O \sin\theta \end{bmatrix} + Sym \left\{ \begin{bmatrix} N_O^T X M_A \Delta_A R_A N_O \sin\theta & N_O^T X M_A \Delta_A R_A N_O \cos\theta \\ -N_O^T X M_A \Delta_A R_A N_O \cos\theta & N_O^T X M_A \Delta_A R_A N_O \sin\theta \end{bmatrix} \right\} < 0.$$

Applying Lemma 4 to the second part in the right side of inequalities in (53), leads to

$$Sym \left\{ \begin{bmatrix} N_C^T M_A \Delta_A R_A Y N_C(\sin\theta)^{-1} & -N_C^T M_A \Delta_A R_A Y N_C(\cos\theta)^{-1} \\ N_C^T M_A \Delta_A R_A Y N_C(\cos\theta)^{-1} & N_C^T M_A \Delta_A R_A Y N_C(\sin\theta)^{-1} \end{bmatrix} \right\}$$
$$= Sym \left\{ \begin{bmatrix} N_C^T M_A & 0 \\ 0 & N_C^T M_A \end{bmatrix} \begin{bmatrix} F_A & 0 \\ 0 & F_A \end{bmatrix} \begin{bmatrix} R_A Y N_C(\sin\theta)^{-1} & -R_A Y N_C(\cos\theta)^{-1} \\ R_A Y N_C(\cos\theta)^{-1} & R_A Y N_C(\sin\theta)^{-1} \end{bmatrix} \right\} \leq$$
$$\eta_1 \begin{bmatrix} N_C^T M_A & 0 \\ 0 & N_C^T M_A \end{bmatrix} \begin{bmatrix} M_A^T N_C & 0 \\ 0 & M_A^T N_C \end{bmatrix} + \eta_1^{-1} \begin{bmatrix} R_A Y N_C(\sin\theta)^{-1} & -R_A Y N_C(\cos\theta)^{-1} \\ R_A Y N_C(\cos\theta)^{-1} & R_A Y N_C(\sin\theta)^{-1} \end{bmatrix}^T \times$$
$$\begin{bmatrix} R_A Y N_C(\sin\theta)^{-1} & -R_A Y N_C(\cos\theta)^{-1} \\ R_A Y N_C(\cos\theta)^{-1} & R_A Y N_C(\sin\theta)^{-1} \end{bmatrix}, \quad (54)$$

$$Sym \left\{ \begin{bmatrix} N_O^T X M_A \Delta_A R_A N_O \sin\theta & N_O^T X M_A \Delta_A R_A N_O \cos\theta \\ -N_O^T X M_A \Delta_A R_A N_O \cos\theta & N_O^T X M_A \Delta_A R_A N_O \sin\theta \end{bmatrix} \right\} =$$
$$Sym \left\{ \begin{bmatrix} N_O^T X M_A & 0 \\ 0 & N_O^T X M_A \end{bmatrix} \begin{bmatrix} F_A & 0 \\ 0 & F_A \end{bmatrix} \begin{bmatrix} R_A N_O \sin\theta & R_A N_O \cos\theta \\ -R_A N_O \cos\theta & R_A N_O \sin\theta \end{bmatrix} \right\} \leq \eta_2 \begin{bmatrix} N_O^T X M_A & 0 \\ 0 & N_O^T X M_A \end{bmatrix} \begin{bmatrix} M_A^T X M_A & 0 \\ 0 & M_A^T X M_A \end{bmatrix} +$$
$$\eta_2^{-1} \begin{bmatrix} R_A N_O \sin\theta & R_A N_O \cos\theta \\ -R_A N_O \cos\theta & R_A N_O \sin\theta \end{bmatrix}^T \begin{bmatrix} R_A N_O \sin\theta & R_A N_O \cos\theta \\ -R_A N_O \cos\theta & R_A N_O \sin\theta \end{bmatrix}.$$

Substituting (54) into (53) and taking the Schur complement of the resultant inequalities we get following inequalities

$$\begin{bmatrix} \lambda_i + \eta_i M_i M_i^T & R_i^T \\ \bullet & -\eta_i I \end{bmatrix} < 0, i = 1,2, \tag{55}$$

which is equivalent to the first inequality of (32) with $\lambda_i$, $M_i$, and $R_i$ defined in (33) for $i = 1,2$. Moreover, inequality (37) with parameters in (36) can be rewritten by considering the uncertainties in (2)-(4).

$$\begin{aligned}
H_{X_{Cl}} + Q^T \widetilde{K}^T P_{X_{Cl}} + P_{X_{Cl}}^T \widetilde{K} Q &= \begin{bmatrix} (\mathbb{A}_0^T X_{Cl} + X_{Cl} \mathbb{A}_0) \sin\theta & (X_{Cl} \mathbb{A}_0 - \mathbb{A}_0^T X_{Cl}) \cos\theta \\ (\mathbb{A}_0^T X_{Cl} - X_{Cl} \mathbb{A}_0) \cos\theta & (\mathbb{A}_0^T X_{Cl} + X_{Cl} \mathbb{A}_0) \sin\theta \end{bmatrix} + \\
&\quad Sym\left\{ \begin{bmatrix} X_{Cl} & 0 \\ 0 & X_{Cl} \end{bmatrix} \begin{bmatrix} M_A & 0 & 0 & 0 \\ 0 & 0 & 0 & 0 \\ 0 & 0 & M_A & 0 \\ 0 & 0 & 0 & 0 \end{bmatrix} \begin{bmatrix} F_A & 0 & 0 & 0 \\ 0 & 0 & 0 & 0 \\ 0 & 0 & F_A & 0 \\ 0 & 0 & 0 & 0 \end{bmatrix} \begin{bmatrix} R_A \sin\theta & 0 & R_A \cos\theta & 0 \\ 0 & 0 & 0 & 0 \\ -R_A \cos\theta & 0 & R_A \sin\theta & 0 \\ 0 & 0 & 0 & 0 \end{bmatrix} \right\} \\
&\quad + Sym\left\{ \begin{bmatrix} X_{Cl} \mathbb{B}_0 \sin\theta & X_{Cl} \mathbb{B}_0 \cos\theta \\ -X_{Cl} \mathbb{B}_0 \cos\theta & X_{Cl} \mathbb{B}_0 \sin\theta \end{bmatrix} \begin{bmatrix} K & 0 \\ 0 & K \end{bmatrix} \begin{bmatrix} \mathbb{C}_0 & 0 \\ 0 & \mathbb{C}_0 \end{bmatrix} \right\} + Sym\left\{ \begin{bmatrix} X_{Cl} & 0 \\ 0 & X_{Cl} \end{bmatrix} \begin{bmatrix} M_B & 0 & 0 & 0 \\ M_B & 0 & 0 & 0 \\ 0 & 0 & M_B & 0 \\ 0 & 0 & M_B & 0 \end{bmatrix} \right. \\
&\quad \begin{bmatrix} F_B & 0 & 0 & 0 \\ 0 & 0 & 0 & 0 \\ 0 & 0 & F_B & 0 \\ 0 & 0 & 0 & 0 \end{bmatrix} \begin{bmatrix} 0 & R_B \sin\theta & 0 & R_B \sin\theta \\ 0 & 0 & 0 & 0 \\ 0 & -R_B \cos\theta & 0 & R_B \cos\theta \\ 0 & 0 & 0 & 0 \end{bmatrix} \begin{bmatrix} K & 0 \\ 0 & K \end{bmatrix} \begin{bmatrix} \mathbb{C}_0 & 0 \\ 0 & \mathbb{C}_0 \end{bmatrix} \right\} + \\
&\quad Sym\left\{ \begin{bmatrix} X_{Cl} \mathbb{B}_0 \sin\theta & X_{Cl} \mathbb{B}_0 \cos\theta \\ -X_{Cl} \mathbb{B}_0 \cos\theta & X_{Cl} \mathbb{B}_0 \sin\theta \end{bmatrix} \begin{bmatrix} K & 0 \\ 0 & K \end{bmatrix} \begin{bmatrix} 0 & 0 & 0 & 0 \\ M_C & 0 & 0 & 0 \\ 0 & 0 & 0 & 0 \\ 0 & 0 & M_C & 0 \end{bmatrix} \begin{bmatrix} F_C & 0 & 0 & 0 \\ 0 & 0 & 0 & 0 \\ 0 & 0 & F_C & 0 \\ 0 & 0 & 0 & 0 \end{bmatrix} \times \right. \\
&\quad \left. \begin{bmatrix} R_C & 0 & 0 & 0 \\ 0 & 0 & 0 & 0 \\ 0 & 0 & R_C & 0 \\ 0 & 0 & 0 & 0 \end{bmatrix} \right\} + Sym\left\{ \begin{bmatrix} X_{Cl} & 0 \\ 0 & X_{Cl} \end{bmatrix} \begin{bmatrix} M_B & 0 & 0 & 0 \\ M_B & 0 & 0 & 0 \\ 0 & 0 & M_B & 0 \\ 0 & 0 & M_B & 0 \end{bmatrix} \begin{bmatrix} F_B & 0 & 0 & 0 \\ 0 & 0 & 0 & 0 \\ 0 & 0 & F_B & 0 \\ 0 & 0 & 0 & 0 \end{bmatrix} \times \right. \\
&\quad \left. \begin{bmatrix} 0 & R_B \sin\theta & 0 & R_B \cos\theta \\ 0 & 0 & 0 & 0 \\ 0 & -R_B \cos\theta & 0 & R_B \sin\theta \\ 0 & 0 & 0 & 0 \end{bmatrix} \begin{bmatrix} K & 0 \\ 0 & K \end{bmatrix} \begin{bmatrix} 0 & 0 & 0 & 0 \\ M_C & 0 & 0 & 0 \\ 0 & 0 & 0 & 0 \\ 0 & 0 & M_C & 0 \end{bmatrix} \begin{bmatrix} F_C & 0 & 0 & 0 \\ 0 & 0 & 0 & 0 \\ 0 & 0 & F_C & 0 \\ 0 & 0 & 0 & 0 \end{bmatrix} \begin{bmatrix} R_C & 0 & 0 & 0 \\ 0 & 0 & 0 & 0 \\ 0 & 0 & R_C & 0 \\ 0 & 0 & 0 & 0 \end{bmatrix} \right\} \le 0.
\end{aligned} \tag{56}$$

Applying Lemma 4 to the second part in the right side of inequalities in (56) and repeatedly taking the Schur complement of the resultant inequalities we get following inequality

$$H_{X_{Cl}} + Q^T \widetilde{K}^T P_{X_{Cl}} + P_{X_{Cl}}^T \widetilde{K} Q \le \begin{bmatrix} \sigma_3 + \sum_{i=3}^{7} \eta_i M_i M_i^T & R_3^T & R_4^T & R_5^T & R_6^T & R_7^T \\ \bullet & -\eta_3 I & 0 & 0 & 0 & 0 \\ \bullet & \bullet & -\eta_4 I & 0 & 0 & 0 \\ \bullet & \bullet & \bullet & -\eta_5 I & 0 & 0 \\ \bullet & \bullet & \bullet & \bullet & -\eta_6 I & 0 \\ \bullet & \bullet & \bullet & \bullet & \bullet & -\eta_7 I \end{bmatrix}, \tag{57}$$

with $\sigma_3$, $M_i$, and $R_i$ defined in (33) for i = 3,...,7. Thus, if the right-hand side of (57) is negative definite, then the constraint (37) is satisfied, which is equivalent to the validity of the third inequality of (32), and this ends the proof. ∎

## 4. Simulation results

In this section, some numerical examples are given to demonstrate the applicability of the proposed method. In this paper, we use YALMIP parser [26] and SeDuMi [27] solver in Matlab tool [28] in order to assess the feasibility of the proposed constraints to obtain the controller parameters.

Dynamic output feedback stabilization problem of the interval fractional-order system (1) is considered with $\alpha = 1.2$ and $A \in A_I = [\underline{A}, \overline{A}]$, $B \in B_I = [\underline{B}, \overline{B}]$, and $C \in C_I = [\underline{C}, \overline{C}]$, where

$$\underline{A} = \begin{bmatrix} -0.9 & -1 \\ 0.8 & -2.6 \end{bmatrix}, \overline{A} = \begin{bmatrix} 1.2 & -2 \\ -1 & -3 \end{bmatrix}, \underline{B} = \begin{bmatrix} 1 \\ 0.9 \end{bmatrix}, \overline{B} = \begin{bmatrix} 1.1 \\ 1 \end{bmatrix}, \underline{C} = [0 \quad -1], \overline{C} = [0 \quad -1]. \tag{58}$$

According to Theorem 1, we can conclude that the uncertain fractional-order system (1), with the parameters in (45), is asymptotically stabilizable using the obtained dynamic output feedback controllers of arbitrary orders, in the form of (29), proposed in Table 1. The location of eigenvalues of the uncertain open-loop system (58) and closed-loop system via obtained output feedback controller with $n_C = 2$ are depicted in Fig. 1. The time response of the resulted uncertain closed-loop FO-LTI system of form (30) with a random system of (58) via obtained controller with $n_C = 2$ and the obtained controller effort are illustrated in Fig. 2 and Fig. 3 respectively, in which, all the trajectories asymptotically converge to zero. Moreover, unlike state feedback procedure, all components of the state vector are not needed to be measured according to the matrix $C$.

Table 1 Controller parameters obtained by Theorem 1.

| $n_c$ | $A_c$ | $B_c$ | $C_c$ | $D_c$ |
|---|---|---|---|---|
| 2 | $\begin{bmatrix} -6 & 6 \\ -7.2083 & -7 \end{bmatrix}$ | $\begin{bmatrix} 1 \\ -0.3975 \end{bmatrix}$ | $\begin{bmatrix} -19.75 \\ 1 \end{bmatrix}^T$ | 1 |

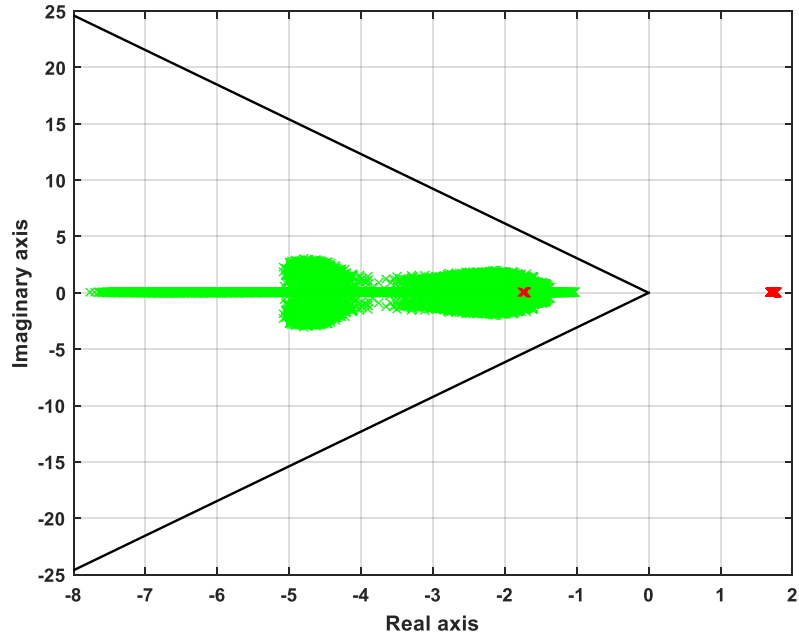

Fig. 1. The location of eigenvalues of the uncertain open-loop system (red) and closed-loop system via obtained output feedback controller with $n_c = 2$ (green).

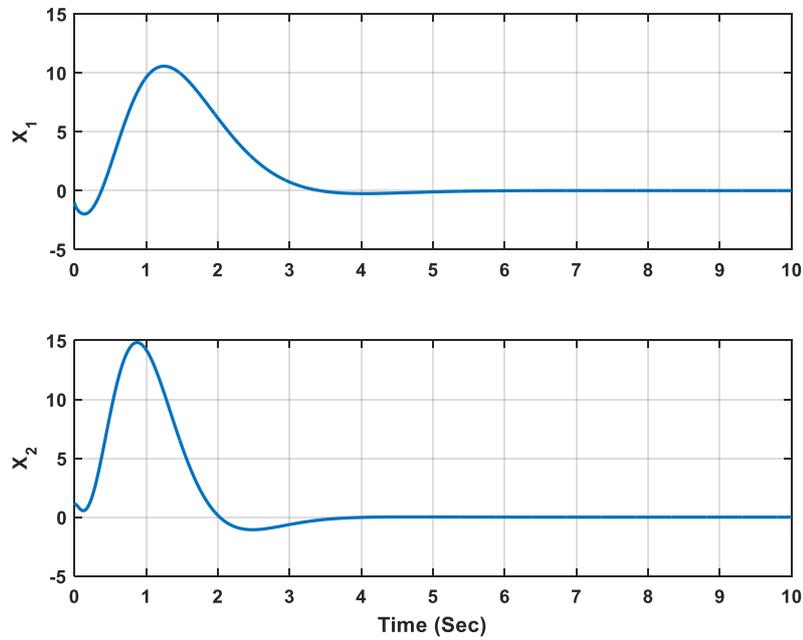

Fig. 2. The time response of the closed-loop system in Example 1 via obtained output feedback controller with $n_c = 2$.

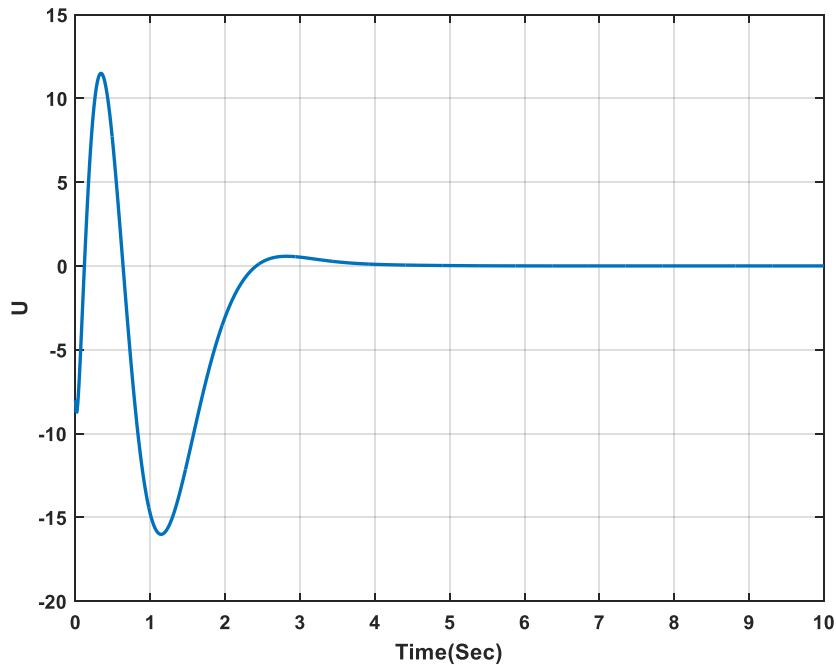

Fig. 3. The control effort of the obtained output feedback controller with $n_c = 2$ in Example 1.

## 5. Conclusion

In this paper the problem of robust dynamic output feedback stabilization of interval FO-LTI systems with the fractional order $1 \leq \alpha < 2$ was solved, in which, a new formulation based on the null-space analysis of the system matrices was proposed in terms of LMIs. The applied uncertain interval model is the most complete model of linear interval systems, in which all of the systems matrices are interval matrices. Sufficient conditions were obtained for designing a robust dynamic output feedback controller that asymptotically stabilizes the interval FO-LTI system, where no limiting constraint is assumed on the state space matrices of the uncertain system. The LMI-based approach of developing robust stabilizing control is preserved in spite of the complexity of assuming the most complete model of uncertain system, including interval output matrix $C$ and the most complete model of linear controller, including direct feedthrough parameter. Eventually, a numerical example with simulations is presented to show the effectiveness of the proposed controller design.